\documentclass[a4paper]{article}
\usepackage{amsfonts}
\usepackage{amssymb}
\usepackage{mathrsfs}
\usepackage{indentfirst}
\usepackage{amsmath}
\usepackage{amsfonts,amssymb}
\usepackage{graphicx,color}
\newcommand{\be}{\begin{equation}}
\newcommand{\ee}{\end{equation}}

\newcommand{\bea}{\begin{aligned}}\newcommand{\eea}{\end{aligned}}

\newcommand{\bec}{\begin{cases}}\newcommand{\eec}{\end{cases}}
\newcommand{\bee}{\begin{eqnarray}}
\newcommand{\eed}{\end{eqnarray}}

\newcommand{\beq}{\begin{eqnarray}}
\newcommand{\edq}{\end{eqnarray}}
\catcode`\@=11 

\catcode`\@=12

\newtheorem{Theorem}{Theorem}[section]

\newtheorem{Lemma}{Lemma}[section]

\newtheorem{Remark}{Remark}[section]
\newtheorem{Property}{Property}[section]

\def\2{{I \hskip -1.0mm I}}
\def\3{{I \hskip -1.0mm I\hskip -1.0mm I}}
\def\4{{I \hskip -0.9mm V}}
\def\6{{V \hskip -1.35mm I}}

\title{Harmonic coordinates in the string and membrane equations}
\author{Chun-Lei
He \;\;and\;\;Shou-Jun
Huang \footnote{Corresponding author: sjhuang@mail.ahnu.edu.cn}
\\ Department of Mathematics, Anhui Normal
University\\ Wuhu 241000, China}\date{ }
\begin{document}
\maketitle
\begin{abstract}In this note, we first show that the solutions to Cauchy problems for two versions of relativistic string and membrane equations are diffeomorphic.  Then we investigate the coordinates transformation presented in Ref. \cite{kongz} (see (2.20) in Ref. \cite{kongz}) which plays an important role in the study on the dynamics of the motion of string in Minkowski space. This kind of transformed coordinates are harmonic coordinates, and the nonlinear relativistic string equations can be straightforwardly simplified into linear wave equations under this transformation.

 \vskip 6mm

 \vskip 3mm

\end{abstract}
\newpage
\section{Introduction}
In recent years the string and membrane theory have drawn great interest (e.g. see \cite{plj}-\cite{hans}). The reason lies in that this theory is not only a possible unification model but also it tightly relates to extremal sub-manifolds in physical space-time from the mathematical point of view.

We recall that the basic equations governing the motion of a $p$-dimensional extended object in a general enveloping space-time $(\mathscr{N},\tilde{g})$ read (cf. \cite{hek})
\begin{equation}\label{1}g^{\mu\nu}\left(x^C_{\mu\nu}-\Gamma^{\rho}_{\mu\nu} x^C_{\rho}+\tilde{\Gamma}^C_{AB}x^A_{\mu}x^B_{\nu}\right)=0,\quad(C=0,1,\cdots,n)\end{equation}
where $x^A_{\mu}=\frac{\partial x^A}{\partial \theta^{\mu}},x^A_{\mu\nu}=\frac{\partial^2 x^A}{\partial \theta^{\mu}\partial\theta^{\nu}}$, $\mu,\nu,\rho=0, 1, ..., p,~~A, B, C=0,1,\cdots, n$.  $g^{\mu\nu}$ is the inverse
of the induced metric $g_{\mu\nu}=\tilde{g}_{\mu\nu}x^A_{\mu}x^B_{\nu}$ on the sub-manifold
$\mathcal{M}$  and $g=-\text{det}(g_{\mu\nu})$. The coordinates
$x^{C}=x^{C}(\theta^0,...,\theta^p)$ describe the sub-manifold
$\mathcal{M}$ and $\tilde{\Gamma}^C_{AB},\Gamma^{\rho}_{\mu\nu}$ stand for the connections of the ambient metric $\tilde{g}$ and the induced metric $g$ respectively. When $p=1$ or $2$, the corresponding equations (\ref{1}) are named as {\it the string equations} or {\it membrane equations}.

For the relativistic string in Minkowski space, it is well known that the equations (\ref{1}) can be simplified by the so-called {\it orthogonal gauge}. Recently based on this gauge, Bellettini {\it et al.} \cite{gjmg} study various properties of closed relativistic strings in Minkowski space. They also show that the convex planar relativistic string with zero initial velocity remains convex in the subsequent times and shrinks to a point while its shape approaches a round circle. However, it is worthy pointing out that the orthogonal gauge adopted in \cite{gjmg} evolves with  time and  in general can not  be solved.

In the papers \cite{kongz}-\cite{hek} , Kong {\it et al.} study the dynamics of relativistic string in Minkowski space and Schwarzschild space-time and the governing equations generally read
\begin{equation}\label{2}g^{\mu\nu}\left(x^C_{\mu\nu}+\tilde{\Gamma}^C_{AB}x^A_{\mu}x^B_{\nu}\right)=0.\quad(C=0,1,\cdots,n)
\end{equation}
In this note we denote $\theta^0=t$ and the Cauchy data associated with (\ref{1}) or (\ref{2}) are  given as
\begin{equation}\label{3}t=0: \quad x=p(\theta), \;\;\;x_t=q(\theta),\end{equation}
where $x=(x^0(\theta^0,\cdots,\theta^p),\cdots,x^n(\theta^0,\cdots,\theta^p))^T.$

 Kong {\it et al.} show that the system (\ref{2})
enjoys some interesting geometric properties. Based on these, they give a sufficient and necessary
condition for the global existence of extremal surfaces without space-like point in $\mathbb{R}^{1+n}$ with given
initial data.  Moreover, numerical analysis show that various topological singularities
develop in finite time in the motion of the string. Surprisingly, they
obtain a general solution formula for this kind of coupled nonlinear equations. With the aid of the
solution formula, they successfully show that the motion of closed strings is always time-periodic.

For the relativistic string in Schwarzschild space-time, under suitable assumptions, He and Kong \cite{hek} most recently obtain the global existence of
smooth solutions of the Cauchy problem (\ref{2})-(\ref{3}).

In this note, for the relativistic string and membrane in a general Lorentzian space-time,  we first show that the solutions to the systems (\ref{1}) and (\ref{2}) with the same Cauchy data (\ref{3}) are  diffeomorphic. In other words, in order to investigate the dynamics of relativistic string and membrane, we only need to consider the global existence of smooth solutions for Cauchy problem (\ref{2})-(\ref{3}). The coordinates transformation adopted in Kong and Zhang \cite{kongz} (see (2.20) in Ref. \cite{kongz}) plays an important role in the study on the dynamics of the motion of relativistic string. we shall show that this kind of coordinates is harmonic \cite{ad} and based on which the nonlinear relativistic string equations in Minkowski space can be straightforwardly simplified to linear wave equations.

\section{Main results}
In this section, we first discuss the relations between the solutions to the Cauchy problems (\ref{1})-(\ref{3}) and (\ref{2})-(\ref{3}). For the convenience of the following discussion, we denote the left hand side of (\ref{2}) by $E$ and \begin{equation*}M=I-G\tilde{g},\end{equation*} where $I$ is the $(1+n)\times(1+n)$ identity matrix,  $G=(G_{AB})_{(1+n)\times(1+n)}\triangleq Qg^{-1}Q^T$, in which
$Q=(X_0,X_1,\cdots,X_p)$ and $X_{\mu}=(x^0_{\mu},x^1_{\mu},\cdots,x^n_{\mu})^T,\;(\mu=0,1,\cdots,p).$

The following lemma comes from He and Kong \cite{hek} (see also \cite{hk}).
\begin{Lemma}The equations (\ref{1}) can be equivalently rewritten as $ME$=0. Hence, any solution to (\ref{2}) which describes the  motion of relativistic string and membrane also satisfies the equations (\ref{1}).\end{Lemma}

\begin{Remark}We would like to point out that the matrix $M$ in fact is the orthogonal projection on the subspace orthogonal to the tangent space of
the sub-manifold  $\mathcal{M}$.\end{Remark}

It is observed that the equations (\ref{1}) can be reduced to (\ref{2}) whenever  the following harmonic conditions hold
\begin{equation}\label{4}\Gamma^{\rho}\triangleq g^{\mu\nu}\Gamma^{\rho}_{\mu\nu}=0,\end{equation}
which are equivalent to require the coordinates $\theta^{\mu}$ are harmonic
\begin{equation}\label{5}\square_g\theta^{\mu}\triangleq g^{\rho\nu}\left(\theta^{\mu}_{\rho\nu}-\Gamma^{\lambda}_{\rho\nu} \theta^{\mu}_{\lambda}\right)=0.\end{equation}
In general, since the connection $\Gamma_{\mu\nu}^{\rho}$ of the induced metric $g$ contains the second order derivatives of the unknown functions $x^C$, it is not easy to solve the equations (\ref{5}).

Aurilia and Christodoulou \cite{ad} firstly apply the harmonic conditions (\ref{4}) or (\ref{5}) to consider the motion of relativistic string and membrane in Minkowski ambient space-time and show that the solutions to (\ref{1}) and (\ref{2}) with given initial data are diffeomorphic. Next we shall derive a similar result as the counterpart in the case that
 the enveloping manifold is a general Lorentzian space-time $(\mathscr{N}, \tilde{g})$.
\begin{Theorem}The solutions to (\ref{1}) and (\ref{2}) with the same Cauchy data (\ref{3}) are diffeomorphic. \end{Theorem}
{\bf Proof.} The  proof  is essentially based on Theorem 2 due to Aurilia and Christodoulou \cite{ad}. Provided with given Cauchy data (\ref{3}), we assume $x_1$ is the solution to Cauchy problem (\ref{2})-(\ref{3}). By Lemma 2.1  this solution  satisfies $(\ref{1})$ and also the same Cauchy data.   On the other hand, if $x_2$ is another solution to (\ref{1}) with the above Cauchy data, then utilizing Theorem 2 in \cite{ad} there subsists a diffeomorphism $f$ such that $x_2=x_1\circ f$. That is to say, the two solutions $x_1$ and $x_2$ are diffemorphic.
Thus, the proof is completed. $\quad\square$

By the results in \cite{kzz}-\cite{hek}, if we impose suitable conditions on Cauchy data, then the system (\ref{2}) has a unique global smooth solution. Then based on Theorem 2.1 this solution  satisfies (\ref{1})-(\ref{3}) uniquely in the sense of diffeomorphism. Since the local coordinates for this solution can be chosen in initial data, hence the solution possesses more apparent geometric meaning and can be easily used to  describe the motion of relativistic string even in a general enveloping Lorentzian space-time $(\mathscr{N}, \tilde{g})$. However, for the relativistic membrane, the global existence of smooth solutions to the Cauchy problem (\ref{2})-(\ref{3})
 still remains open since the difficulty arises in that this hyperbolic system is multidimensional and highly nonlinear.

In the sequel, we show that for the relativistic string moving in Minkowski space,  the system (\ref{2}) can be directly simplified to
linear wave equations with the coordinates transformation introduced by Kong and Zhang \cite{kongz}.

Let  \begin{equation}\label{7}\lambda_{\pm}=\frac{-g_{01}\pm\sqrt{g_{01}^2-g_{00}g_{11}}}{g_{11}},\end{equation}
where $g_{00}=|x_t|^2-1,g_{01}=g_{10}=\langle x_t,x_{\theta}\rangle,g_{11}=|x_{\theta}|^2$.

The following equations play an important role in the study on global existence of smooth solutions for the Cauchy problem (\ref{2})-(\ref{3})
\begin{equation}\label{8}\frac{\partial\lambda_{\pm}}{\partial t}+\lambda_{\mp}\frac{\partial\lambda_{\pm}}{\partial \theta}=0.\end{equation}
The corresponding  initial data  take the following form respectively,
\begin{equation}\label{9}t=0:\quad \lambda_{\pm}=\Lambda_{\pm}(\theta),\end{equation}
where
\begin{equation}\label{10}\Lambda_{\pm}(\theta)=\frac{1}{|p'(\theta)|^2}\left(-\langle q(\theta),p'(\theta)\rangle\pm\sqrt{\langle q(\theta),p'(\theta)\rangle^2-(|q(\theta)|^2-1)|p'(\theta)|^2}\right)\end{equation}
\begin{Remark}We would like to point out that for the string case, the equations (\ref{8}) are valid no matter whether the enveloping space-time is curved or not. Moreover, it is easy to verify that by constructing some geometric invariants, the nonlinear
equations (\ref{2}) can be converted into a set of semilinear wave equations ( e.g. see \cite{hek}). Nevertheless, this situation for membrane does not hold.\end{Remark}

Recall the coordinates transformation in Kong and Zhang \cite{kongz}
\begin{equation}\label{11}(t,\theta)\rightarrow(t,\sigma): \quad \theta=\Theta(t,\sigma), \end{equation}
where
\begin{equation}\label{12}\Theta(t,\sigma)=\frac12\int_0^{\sigma+t}\Lambda_+(\varrho(\xi))d\xi-\frac12\int_0^{\sigma-t}\Lambda_-(\varrho(\xi))d\xi,\end{equation}
in which $\theta=\varrho(\sigma)$ denotes the inverse function of
\begin{equation}\label{13}\sigma=\rho(\theta)=\int_0^{\theta}\frac{2}{\Lambda_+(\xi)-\Lambda_-(\xi)}d\xi.\end{equation}
Let $\sigma=\Phi(t,\theta)$ be the inverse function of $\theta=\Theta(t,\sigma)$.
Then the solution of the Cauchy problem (\ref{8}) and (\ref{9}) can be expressed as
\begin{equation}\label{14}\lambda_{\pm}(t,\theta)=\Lambda_{\pm}(\varrho(\Phi(t,\theta)\pm t)).\end{equation}
\begin{Remark}For simplicity, the new coordinates $(t,\sigma)$ are called Kong-Zhang coordinates which play an important role in
our future study.\end{Remark}

Under some reasonable assumptions, the mapping defined by (\ref{11})-(\ref{12}) is globally diffeomorphic (see \cite{kongz}-\cite{hek} for details). We now turn to simplify the equations (\ref{2}) under the Kong-Zhang coordinates.

Noting (\ref{14}), by a direct calculation we have
\begin{eqnarray}\label{15}\frac{\partial \Theta}{\partial t}=-\frac{g_{01}}{g_{11}}=-\frac{\langle x_t,x_{\theta}\rangle}{|x_{\theta}|^2},\end{eqnarray}
\begin{eqnarray}\label{16}\frac{\partial \Theta}{\partial \sigma}=\frac{\sqrt{g_{01}^2-g_{00}g_{11}}}{g_{11}}=\frac{1}{|x_{\theta}|^2}\sqrt{\langle x_t,x_{\theta}\rangle^2-(|x_t|^2-1)|x_{\theta}|^2}.\end{eqnarray}
It obtains from  (\ref{15})-(\ref{16}) that
\begin{eqnarray}\label{17}\frac{\partial^2\Theta}{\partial t^2}-\frac{\partial^2\Theta}{\partial \sigma^2}=0.\end{eqnarray}
Denote $$\tilde{x}(t,\sigma)=x(t,\Theta(t,\sigma)).$$
By direct computation, we have
$$\tilde{x}_t=x_t+x_{\theta}\frac{\partial \Theta}{\partial t}, \quad \tilde{x}_{\sigma}=x_{\theta}\frac{\partial \Theta}{\partial \sigma}$$
and
\begin{equation}\label{18}\tilde{x}_{tt}=x_{tt}+x_{t\theta}\frac{\partial \Theta}{\partial t}+\left(x_{\theta t}+x_{\theta\theta}\frac{\partial\Theta}{\partial t}\right)\frac{\partial\Theta}{\partial t}+x_{\theta}\frac{\partial^2\Theta}{\partial t^2},\end{equation}
\begin{equation}\label{19}\tilde{x}_{\sigma\sigma}=x_{\theta\theta}\left(\frac{\partial\Theta}{\partial \sigma}\right)^2+x_{\theta}\frac{\partial^2\Theta}{\partial \sigma^2}.\end{equation}
Thus, combining the equations (\ref{15})-(\ref{19}), the equations (\ref{2}) can be converted into the following linear wave equations
\begin{equation}\label{20}\tilde{x}_{tt}-\tilde{x}_{\sigma\sigma}=0.\end{equation}
The initial data  (\ref{3}) are transformed to
\begin{equation}\label{21}t=0: \tilde{x}=p(\varrho(\sigma)),\quad \tilde{x}_t=q(\varrho(\sigma)).\end{equation}
Noting Remark 2.2 and summarizing the above discussions leads to
\begin{Theorem}Under the Kong-Zhang coordinates, the equations (\ref{2}) governing the motion of relativistic string in Minkowski or Lorentzian ambient space-time can be directly simplified into linear wave equations or semilinear wave equations respectively.
\end{Theorem}
Moreover we have
\begin{Property}The coordinates transformation defined by (\ref{11})-(\ref{12}) satisfies the following constraints
\begin{equation}\label{22}\langle\tilde{x}_t,\tilde{x}_{\sigma}\rangle=0,\quad |\tilde{x}_t|^2+|\tilde{x}_{\sigma}|^2=1.\end{equation}\end{Property}
{\bf Proof.}  Noting (\ref{15}) gives
\begin{equation}\label{23}\langle\tilde{x}_t,\tilde{x}_{\sigma}\rangle=\left\langle x_t+x_{\theta}\frac{\partial\Theta}{\partial t},x_{\theta}\frac{\partial\Theta}{\partial \sigma}\right\rangle
=\frac{\partial\Theta}{\partial \sigma}\left(\langle x_t,x_{\theta}\rangle+|x_{\theta}|^2\frac{\partial\Theta}{\partial t}\right)=0.\end{equation}
\begin{equation}\label{24}|\tilde{x}_t|^2+|\tilde{x}_{\sigma}|^2=|x_t|^2+|x_{\theta}|^2\left(\frac{\partial\Theta}{\partial t}\right)^2+2\langle x_t,x_{\theta}\rangle\frac{\partial\Theta}{\partial t}+x^2_{\sigma}\left(\frac{\partial\Theta}{\partial\sigma}\right)^2=1.\end{equation}
In the last equality of (\ref{24}), we have made use of the equations (\ref{15}) and (\ref{16}). The proof is completed. $\quad\square$

\begin{Remark}As shown in Remark 2.2,  if the ambient space-time is curved, the equations (\ref{8}) are still valid. Thus, by the above method we can  prove the Kong-Zhang coordinates also satisfy  (\ref{22}).\end{Remark}
From Property 2.1, we can see that the coordinates transformation (\ref{11})-(\ref{12})  is one type of orthogonal gauge (cf. \cite{gjmg}). Moreover, this coordinates transformation  actually satisfies the harmonic conditions (\ref{4}) or (\ref{5}).

\section{Concluding remarks}
In this note we first obtain that in order to study the dynamics of relativistic string and membrane, it is equivalent to consider the
global smooth solutions of the Cauchy problem (\ref{2})-(\ref{3}). This result can be viewed as an extension of \cite{ad}. Actually, the model equations (\ref{2}) for the membrane in Minkowski space are established in \cite{hk} and their many interesting geometric properties have been
explored. The global smooth solutions to the Cauchy problem (\ref{2})-(\ref{3}) is very worth investigating in the future.

It is well-known that the orthogonal gauge is very important in the string theory. But due to the fact that this gauge condition is a kind of nonlinear  constraints, so in general it can not be solved.  The second purpose of this note is to emphasize that the coordinates transformation (\ref{11})-(\ref{12}) given by Kong and Zhang gives us one kind of such solution and moreover it is globally diffeomorphic. On one hand, the Kong-Zhang coordinates are determined through the Cauchy data, therefore the parameters in the sub-manifold can possess better geometric meaning.
For these reasons, we believe that, by Kong-Zhang coordinates, we can obtain the global existence of smooth solutions to
the Cauchy problem (\ref{2})-(\ref{3}) without any small condition assumed on the initial data (cf. \cite{hek}).  On the other hand, along this line it is interesting to consider whether there subsists such coordinates transformation for the membrane equations.

\vskip 5mm\noindent{\Large {\bf Acknowledgements.}} This work was supported
by the Foundation for University's Excellent Youth Scholars (Grand Nos. 2009SQRZ025ZD, 2010SQRL025) and the
University's Natural Science Foundation from Anhui Educational Committee (Grand No. KJ2010A130).

 \end{document}